\documentclass[comsoc,12pt]{IEEEtran}
\usepackage{amsmath,amsfonts}
\usepackage{algorithmic}
\usepackage{algorithm}
\usepackage{array}
\usepackage[caption=false]{subfig}
\usepackage{textcomp}
\usepackage{stfloats}
\usepackage{url}
\usepackage{verbatim}
\usepackage{graphicx}
\usepackage{cite}
\usepackage[most]{tcolorbox}

\usepackage{booktabs}     
\usepackage{caption}      
\usepackage{cleveref}     
\usepackage{makecell}
\usepackage{pifont} 
\newcommand{\cmark}{\ding{51}}%
\newcommand{\xmark}{\ding{55}}%
\usepackage{multirow}
\usepackage{subcaption}
\usepackage{xspace}

\usepackage{eso-pic}

\newcommand{\sysname}{Holoscope}
\newif\ifsubmission
\submissiontrue

\ifsubmission
    \newcommand{\as}[1]{}
    \newcommand{\mm}[1]{}
    \newcommand{\id}[1]{}
    \newcommand{\todo}[1]{}

\else
    \newcommand{\as}[1]{\textit{\color{cyan}[Andrea: #1]}}
    \newcommand{\mm}[1]{\textit{\color{purple}[Marco: #1]}}
    \newcommand{\id}[1]{\textit{\color{green}[Idilio: #1]}}
    
\fi

\begin{document}
\crefname{figure}{Fig.}{Figs.}
\Crefname{figure}{Fig.}{Figs.}

\crefname{tabular}{Table}{Tables}
\Crefname{tabular}{Table}{Tables}

\crefname{section}{Sec.}{Secs.}
\Crefname{section}{Sec.}{Secs.}
\title{Holoscope: Open and Lightweight\\ Telescope \& Honeypot Platform}

\author{
\IEEEauthorblockN{
Andrea Sordello, Marco Mellia, Idilio Drago, Rodolfo Valentim, Francesco Musumeci
}
\IEEEauthorblockN{
Massimo Tornatore, Federico Cerutti, Martino Trevisan, Alessio Botta, Willen B. Coelho
}}%



\IEEEpubid{This work is licensed under a Creative Commons Attribution 4.0 License. For more information, see https://creativecommons.org/licenses/by/4.0}

\maketitle
\AddToShipoutPictureFG*{%
  \AtPageUpperLeft{%
    \raisebox{-1.2cm}{%
      \hspace{1.6cm}%
      \parbox{0.8\textwidth}{%
        \footnotesize\itshape
        Accepted for publication in IEEE Communications Magazine\\
        DOI: 10.1109/MCOM.001.2500784
      }%
    }%
  }%
}
\begin{abstract}
The complexity and scale of Internet attacks call for distributed, cooperative observatories capable of monitoring malicious traffic across diverse networks. \sysname{} is an open, lightweight, and cloud-native platform designed to simplify the deployment and management of telescope (passive) and honeypot (active) sensors. 
Built upon K3s and WireGuard, \sysname{} offers secure connectivity, automated sensor onboarding, and resilient operation even in resource-constrained environments. Through modular design and Infrastructure-as-Code principles, it supports dynamic sensor orchestration, automated recovery, and data processing. We build, deploy, and operate \sysname{} across multiple institutions and cloud networks in Europe and Brazil, enabling unified visibility into large-scale attack phenomena while maintaining ease of integration and security compliance.
\end{abstract}
\section{Introduction}

Security monitoring infrastructures play a central role in identifying emerging cyber threats and characterising malicious Internet activity.
Among these, network telescopes (darknets) monitor unused, routable IP address space that hosts no services. Since no legitimate traffic should target these addresses, the unsolicited traffic they receive typically consists of scanning campaigns targeting exposed hosts and services, unveiling potential malicious behaviour such as worm propagation and botnet reconnaissance~\cite{Cloud_Telescope_long}. In contrast, honeypots are decoy systems that emulate vulnerable services, enabling active interaction with attackers and observation of exploitation attempts, command execution, lateral movement, and post-compromise activities.

Darknets and honeypots are often deployed within a single network domain, making the observed traffic strongly dependent on factors such as network prefix size, autonomous system (AS), and geographic location, as malicious activities are often localised on specific address ranges~\cite{scanning_the_scanners}.
 
Achieving a broader and more comprehensive view of malicious behaviour requires combining observations from multiple heterogeneous networks~\cite{all_darknet_are_the_same,MetaTelescope,SweetsPot,pauley_dscope_2023}. However, building such collaborative infrastructure is challenging due to the need for secure data sharing, automation, scalability, and cross-organisational management.
\begin{figure}[]
    \centering\includegraphics[width=\linewidth]{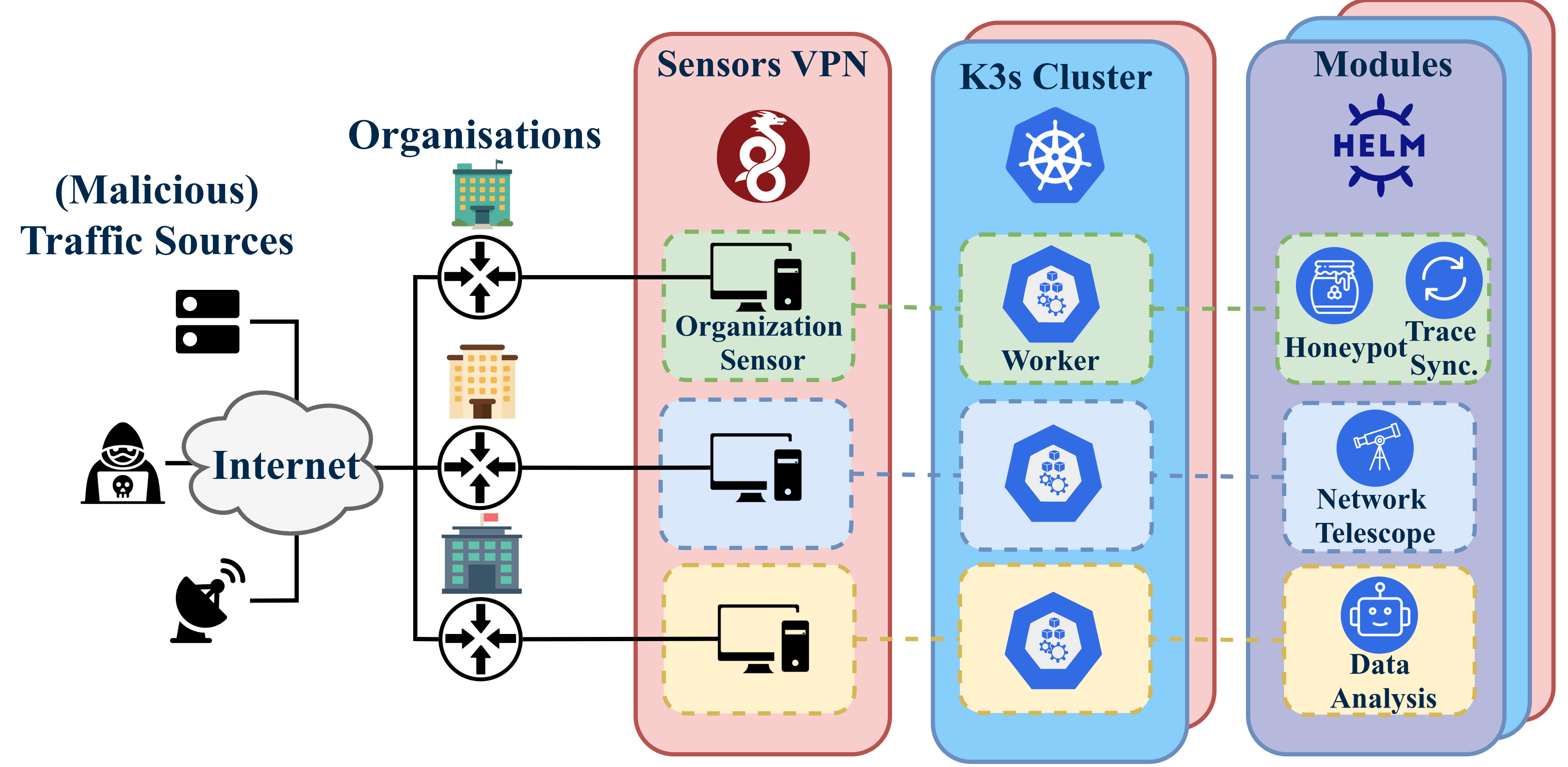}
    \caption{\sysname{} architecture. Each organisation provides a sensor exposed to the Internet. \sysname{} activates modules to collect and analyse malicious traffic.
    }
    \label{fig:holoscope_architecture}
\end{figure}

We introduce \sysname{}, a distributed and scalable cybersecurity observatory developed within the PROTECT-IT/SERICS~\cite{serics2026} project. \sysname{} simplifies the deployment and management of Internet-exposed sensors across geographically distributed networks owned by different organisations. It enables a configurable network of sensors that continuously collect, process, and share unsolicited traffic from diverse vantage points, producing high-value data for modelling attack patterns and studying large-scale adversarial behaviour. We provide an overview of \sysname{} in~\cref{fig:holoscope_architecture}. 
\IEEEpubidadjcol
At the time of writing, \sysname{} is running across 10 organisations — including universities, service providers, and cloud infrastructures — in Europe, the US, and Brazil. It runs on Unix-based physical or virtual systems, ensuring broad deployability. The platform is publicly available at https://github.com/SmartData-Polito/Holoscope. The traffic dataset is available under a non-disclosure agreement due to the sensitive nature of the traffic traces.
The remainder of the paper is organised as follows: \Cref{sec:rel_work} reviews related work on distributed telescopes and honeypot networks; \Cref{sec:holoscope} presents \sysname{}’s design, operation, and modules; \Cref{sec:deployment} describes its real-world deployment and observed traffic insights; and \Cref{sec:conclusion} discusses the lessons learned.

\section{Related Work}
\label{sec:rel_work}
Distributed infrastructures for collecting unsolicited Internet traffic have evolved along two main directions: (i) large-scale darknet platforms~\cite{Cloud_Telescope_long} and (ii) coordinated honeypot systems~\cite{survey_honeypots}. We consider recent solutions that are not deprecated, support replication through distributed sensors, and operate in real-world deployments rather than simulations or single-organisation testbeds. A representative list is provided in~\Cref{tab:rel_work}.

At the national scale, distributed darknet initiatives include the US-based Merit National Distributed Network Telescope (NDNT)~\cite{ndnt_merit} and the Greek EWIS platform~\cite{EWIS}, both of which federate sensors across multiple administrative domains. Unlike \sysname{}, EWIS depends on preconfigured hardware distributed to participants, limiting scalability and flexibility. NDNT focuses on federating existing telescope sensors and IP address space, rather than offering a reproducible open-source deployment framework. Cloud-based distributed telescope designs have also been explored~\cite{Cloud_telescope,pauley_dscope_2023}, demonstrating the feasibility of deploying darknet sensors in cloud environments. \sysname{} likewise supports cloud-hosted sensors, enabling elastic deployment, geographic diversity, and multi-provider scalability.

The honeypot ecosystem has likewise evolved toward distributed architectures. Recent versions of the widely used T-Pot~\cite{telekom_tpotce} support distributed deployments, and other platforms, such as SweetsPot~\cite{SweetsPot}, enable the orchestration of multiple honeypots across geographically and administratively distributed environments. In contrast to \sysname{}, these platforms primarily focus on honeypot deployment, with limited emphasis on sensor configuration and orchestration mechanisms. Other solutions, such as GCA AIDE~\cite{gcaaide}, are built around proprietary honeypot technologies.

A smaller set of platforms integrates both darknet and honeypot monitoring. One example is CAIDA’s iVoyager~\cite{caida_ivoyager}, a novel project that combines a central experimentation node with lightweight distributed sensors. Compared to \sysname{}, which prioritises large-scale darknet data collection and analysis, iVoyager places stronger emphasis on offering a centralised research infrastructure.

\begin{table}[]
\caption{Overview of existing platforms.}
\centering

\resizebox{\linewidth}{!}{
\begin{tabular}{l| c c c c }
\hline
\multirow{2}{*}{\textbf{Platform}} & 
\multirow{2}{*}{\textbf{\makecell{Platform\\of ...}}}
&\multirow{2}{*}{ \textbf{\makecell{3rd-Party \\Sensor}}} &\multirow{2}{*}{\textbf{\makecell{Sensor\\ Config.}}}&
\multirow{2}{*}{ \textbf{\makecell{Experiment\\Set}}}  \\

 \\
\hline

 SweetsPot \cite{SweetsPot} & Honeypot & \cmark & \xmark & Custom \\
  GCA AIDE \cite{gcaaide} & Honeypot  & \cmark & \xmark & Proprietary  \\
  T-POT \cite{telekom_tpotce} & Honeypot  & \cmark & \xmark & T-POT \\

 Merit NDNT\cite{ndnt_merit}& Darknet   & -- & -- &-- \\
 EWIS\cite{EWIS}&Darknet  & \cmark & \cmark & --  \\
 Bortoluzzi et al. \cite{Cloud_telescope} & Darknet & \xmark& \cmark & -- \\
 iVoyager\cite{caida_ivoyager} & Both  & \cmark &\cmark & Custom\\
\hline
\hline 
\multicolumn{1}{c|}{ \textbf{\sysname}} 
& \textbf{\makecell{Both}} 
 & \cmark &\cmark & Custom 

\\[0.8ex]
\hline 
\multicolumn{5}{l}{Legend: \cmark Supported, \xmark Not supported, --  Not applicable/Not provided}
\end{tabular}
}
\label{tab:rel_work}
\end{table}

\section{\sysname{}}
\label{sec:holoscope}

 \sysname{} must satisfy two primary requirements: (i) supporting operation across distributed and heterogeneous resource-constrained sensors and (ii) providing a flexible and responsive execution environment for applications.

\subsection{\sysname{} design}

We now describe the design of \sysname{}’s architectural layers, motivating their purpose and explaining how they interact with one another.
\paragraph{Secure Connectivity Layer.}
The foundation of \sysname{} is a secure layer that interconnects all participating sensors, located across the networks of third-party organisations. \sysname{} relies on a Virtual Private Network (VPN) that provides confidentiality and integrity of communications while normalising networking across heterogeneous environments. This abstraction mitigates issues related to Network Address Translation (NAT), firewalls, and dynamic IP addressing.
The VPN topology follows a point-to-point model between each sensor organisation and the central server(s). This design prevents direct sensor-to-sensor visibility and limits lateral movement in case of compromise, strengthening the overall security posture.

We implement this layer using WireGuard, a lightweight modern VPN solution.
\paragraph{Orchestration layer}

To meet the requirements of flexibility and responsiveness in application management, \sysname{} adopts cloud-native technologies. Docker containers provide an effective mechanism to package and distribute applications across sensors by encapsulating each application together with its dependencies into lightweight, isolated environments. This approach ensures portability across heterogeneous devices and prevents conflicts arising from differing system configurations.

However, distributing containers across multiple geographically dispersed nodes requires automated coordination. For this reason, the orchestration layer leverages a container orchestrator to establish a distributed cluster on top of the secure network. The orchestrator is responsible for scheduling workloads, maintaining their desired state, handling failures, and enforcing isolation and security policies through an off-the-shelf solution. By automating deployment, scaling, and recovery, the orchestration layer enables \sysname{} to dynamically adapt to changes in application requirements and variations in node availability.

We implement the orchestration layer using K3s, a lightweight Kubernetes distribution. The adoption of K3s provides several advantages, including:
\begin{itemize}
    \item efficient operation in resource-constrained environments 
   
    \item application availability and self-healing through Kubernetes primitives 
    \item  redundancy via replicated \textit{etcd} nodes, improving platform resiliency 
\end{itemize}

\paragraph{Module layer}
The top layer of \sysname{} comprises modular applications deployed by the container orchestrator. We refer to these applications as \emph{modules} because each provides a self-contained application that can be independently executed without affecting the rest of the platform. A module includes not only the application container but also the associated Kubernetes resources, e.g., services and storage components, required for its correct operation. Example modules include darknet, honeypots, and traffic analysis tools. We present the complete list in \cref{sec:modules}.

\subsection{\sysname{} operation}

We now describe how \sysname{} manages its sensors, focusing on the processes for onboarding new ones, performing their initial configuration, maintaining and updating deployed sensors, and tracking organisation-specific requirements, such as network interface configurations and IP addressing.

To automate and systematically control these configuration tasks, we adopt Ansible, an open-source automation engine that manages infrastructure through reusable playbooks, i.e., declarative configuration code. This enforces the Infrastructure-as-Code (IaC) paradigm as a core principle of \sysname{}, ensuring reproducibility, consistency, auditability, and scalability across heterogeneous and geographically distributed environments, while reducing manual intervention and configuration drift.

\paragraph{New sensor onboarding}
A fundamental operational task in \sysname{} is the onboarding of new sensors. To enable this process, the sensor must be reachable over the Internet via SSH, which allows the automated execution of the Ansible playbook responsible for this task. This playbook (i) installs the required software components, including Docker, K3s, and WireGuard; (ii) configures the additional virtual interface for the VPN connection that will be used for the sensor's management; and (iii) registers the sensor within the K3s cluster. 

By automating the entire onboarding process, the Ansible-based approach significantly reduces manual intervention by the sensor owner and ensures a consistent, reproducible deployment process. 

\paragraph{Adding modules to \sysname{}}
\sysname{} offers a private container image registry that we use to deploy modules (see Sec. \ref{sec:modules}). The choice to offer a private registry avoids depending on a third-party registry, while keeping our modules secure. To add a new module to \sysname{}, it is sufficient to provide a \textit{Dockerfile}, which can be built and uploaded to the private registry.

\paragraph{Deploying modules on sensors} 

A key aspect of \sysname{} is the heterogeneity of its sensors. Each sensor has its own configuration, e.g., network interface card (NIC) names, IP address ranges as well as specific policies. To handle this heterogeneity, we employ Helm, a Kubernetes package manager that allows creating parametrised, reusable flexible deployments. By using Helm, we define module templates and later customise them with sensor-specific details, thereby extending the Infrastructure-as-Code paradigm also to module deployment.

\subsection{Modules offered by \sysname{}}
\label{sec:modules}

We now present the set of core applications available in \sysname{}. We refer to them as modules, as each operates independently and runs as a self-contained component.

\paragraph{Darknet - Passive telescope}

The Darknet module activates a passive sensor over an unused IP address range owned by the organisation. It supports three deployment configurations:

(i) The organisation statically routes the entire darknet prefix from the ingress router to the sensor’s production interface. This requires a one-time modification of the routing policy.

(ii) The organisation assigns the full darknet address space directly to the sensor’s NIC. This is the simplest setup: unsolicited traffic reaches the sensor as regular traffic, and no responses are generated. Cloud-based darknet deployments commonly adopt this model, treating darknet IPs as standard routable addresses~\cite{Cloud_telescope,pauley_dscope_2023}.

(iii) A custom module performs transparent Layer-2 redirection by replying to ARP requests from ingress routers, associating darknet IPs with the sensor’s MAC address and forwarding traffic accordingly. This approach has been adopted in prior darknet deployments~\cite{Practical_Darknet_Measurement,EWIS}.

In all configurations, the Darknet module coordinates with the Network Enforcement module to install appropriate iptables rules that block any outgoing traffic from the darknet address range.

\paragraph{Honeypot - Active responder}

\sysname{} supports any honeypot that can be containerised. Deploying containerised honeypots within a cluster introduces two main challenges: (i) exposing the pod hosting the honeypot transparently, and (ii) preventing any lateral movement originating from the pod itself, as it runs vulnerable code.

We address the first challenge through the Network Enforcement Module which manages traffic steering and port exposure. To preserve honeypot integrity and minimise the risk of compromise, we rely on Kubernetes' built-in isolation mechanisms, ensuring each pod operates with the minimum required privileges. Since this is a static approach, we additionally integrate Falco, an active monitoring system that tracks runtime behaviour and detects signs of compromise.

\paragraph{Network Enforcement}
Although Kubernetes provides native networking abstractions, such as ClusterIP and NodePort services, these mechanisms are insufficient for darknet and honeypot deployments. Such scenarios require fine-grained traffic steering and enforcement policies beyond standard Kubernetes services.

For instance, we must block any outgoing packets generated from darknet IP addresses to preserve their passive nature. Moreover, the exposure of a honeypot using a sensor port inevitably requires using the Kubernetes \textit{hostNetwork} option. However, this creates a security risk, since the pod would have unrestricted access to the sensor's root network namespace.

Motivated by these limitations, we introduce the novel Network Enforcement Module to obtain a more fine-grained traffic control mechanism.

Its task is to handle unsolicited traffic arriving at the sensor’s NIC and forward it to the appropriate module (e.g., honeypot or darknet), making them appear as if they are running as real services. This redirection is done by inserting proper rules inside the sensor's iptables and executing the module as a privileged DaemonSet. At runtime, the module generates and updates the sensor's iptables based on the modules currently running on the sensor, enforcing different exposure rules dynamically. Thanks to this steering at the kernel level, we limit the overhead.

The use of this module provides additional advantages, including rate limiting—crucial to prevent the sensor from inadvertently participating in Distributed Denial of Service (DDoS) attacks—and the ability to expose a honeypot instance across a range of ports or IP addresses without requiring multiple deployments. For example, \sysname{} can route traffic from attackers to different backend services or steer packets targeting selected address ranges to basic Layer-4 responders that complete the TCP three-way handshake to capture initial payloads~\cite{all_darknet_are_the_same}.

The module prevents interference with the native iptables chains managed by the K3s Container Network Interface (CNI) by using additional custom iptables chains.

\paragraph{Traffic Collector}

This module captures all traffic directed to the monitored address space of each sensor, while excluding management traffic through properly configured capture filters. It uses tcpdump to save packet traces on the sensor's disk.

\paragraph{Log Sync}
The module retrieves raw traffic traces and other logs from sensors and stores them in a centralised storage using Rsync, a data transferring tool. 
Organisations can limit or disable this feature, e.g., in cases where organisation-specific data retention policies apply.

\paragraph{Data Analysis}

Finally, \sysname{} supports modules dedicated to analysing the traffic and logs collected by the sensors. Because these tasks may impose significant computational overhead, we typically execute them on the most powerful sensor available, subject to the hosting organisations' policies.
The analyses range from simple aggregated metric computation to more advanced techniques leveraging machine learning and artificial intelligence (ML/AI).

\section{Deployment of \sysname{}}
\label{sec:deployment}

We now describe our deployment and provide an overview of 3 months of collected traces.

\subsection{Participating organisations}
\label{ss:organisations}
\begin{table}[]
\caption{Organisations contributing to \sysname{}.}
\centering

\resizebox{\linewidth}{!}{
\begin{tabular}{c|c|c c c| c c}
\hline
\multirow{2}{*}{\textbf{Sensor}} & 
\multirow{2}{*}{\textbf{Organisation}} & 
\multirow{2}{*}{\textbf{Country}}&
\multirow{2}{*}{ \textbf{\makecell{Network\\Telescope}}} &
\multirow{2}{*}{ \textbf{\makecell{Honeypot\\Allowed?}}} &
\multicolumn{2}{c}{ \textbf{\makecell{Daily /24 statistics }}} \\
&& & & &  
\textbf{\makecell{\# Packets}} & 
\textbf{\makecell{\# Senders}} \\
\hline
$A_1$, $A_2$ & \textbf{Politecnico di Torino} & ITA & /23 & \cmark  & 4.1M & 45k \\
$B_1$, $B_2$ & Consortium GARR & ITA & /23 & -  & 3.4M & 44k \\
$C$ & Politecnico di Milano & ITA & /23 & \cmark  & 1.3M & 37k \\
$D$ & Università di Trieste & ITA & /22 & -  & 1.9M & 42k \\
$E$ & Università di Brescia & ITA & /24 & \cmark  & 18M & 101k \\

$F_1$ & UFES & BRA & /25 & \cmark  & \multirow{2}{*}{2.7M} & \multirow{2}{*}{42k} \\
$F_2$ & IFES & BRA & /25 & \cmark  &  &  \\
$G$ & RNP & BRA & /19 & - & 6.3M & 42k \\

$H$ & Microsoft Azure & East US & /24 & \cmark & 4.7M & 48k \\
$I$ & Univ. di Napoli Federico II & ITA & - & \cmark & - & --  \\
\hline
\hline 
\multicolumn{2}{c|}{ \textbf{\sysname}} 
& \textbf{\makecell{Global}} 
& \textbf{\makecell{Available \\11\;520 IPs}} 
& \multicolumn{1}{c|}{\textbf{\makecell{Generally \\Allowed}}} &
 
\\[0.8ex]
\hline 
\end{tabular}
}
\label{tab:organisations}
\end{table}
We deploy \sysname{} through collaboration with several Italian and Brazilian organisations, as shown in~\Cref{tab:organisations}. All organisations provide an IP range within their address space to install a darknet. Some support honeypot deployment and workload execution, e.g., data processing, directly within their infrastructure.

Each organisation contributes subnet ranges of varying sizes and sensor capabilities (ranging from 4 GB RAM and 4 vCPU to 32 GB RAM and 16 vCPU). To enable a fair comparison, in our analysis, we consider only a /24 subnet per sensor. Sensors $F_1$ and $F_2$, which correspond to two contiguous /25 networks located in two different campuses, are treated jointly as a single /24 sensor, and denoted as $F$. The sensors joined \sysname{} at different times. For instance, we activated sensor $H$, hosted in Microsoft Azure East US, in August 2025. A single /24 darknet typically observes between 1.3 million and 6.3 million packets per day from approximately 42 thousand unique external senders. 
\subsection{Resource overhead}
We evaluate Holoscope’s computational resource requirements by quantifying the platform overhead relative to standalone experiment execution. To avoid impacting the production deployment described in~\cref{ss:organisations}, we conduct a benchmark by deploying \sysname{} on dedicated VMs (2 vCPUs and 2 GB of RAM each).

Holoscope introduces an overhead of approximately 300 MB of RAM and 3\% to 10\% CPU usage per sensor. The sensor acting as the K3s master node has higher resource requirements, consuming between 0.6 and 1 GB of RAM and around 25\% CPU, due to additional components running on it, e.g., the K3s \textit{etcd} database.

\sysname{} requires only a few GB of disk space, while the volume of downloaded data during the installation is on the order of MB. The management traffic exchanged, over the VPN, by sensors consists predominantly of K3s/VPN management messages.

\subsection{\sysname{} traffic profile}

We analyse traffic collected by \sysname{} sensors from July 2025 to October 2025. During this period, some sensors hosted at the same time both active and passive experiments, we refer to them as ``Responder $X$" and ``Darknet $X$". We examine traffic in terms of: (i) temporal evolution, (ii) sender similarity, and (iii) targeted destination ports.

\subsubsection{Traffic temporal profile} 
\begin{figure*}[t]
\centering
\subfloat[Flows observed by each IP address across different sensors within a 24-hour timeslot.]{
    \includegraphics[width=0.28\textwidth]{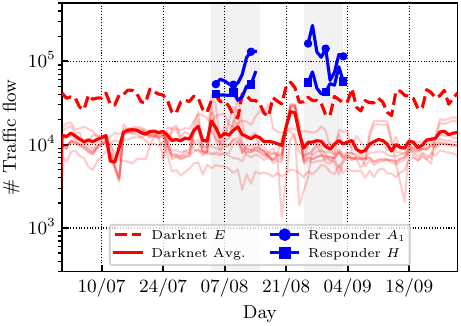}
    \label{fig:timeline_combined}
}
\hfill
\subfloat[Common sender observed from different \slash24 darknet sensors from 25 July to 3 August (10 days).]{
    \includegraphics[width=0.28\textwidth]{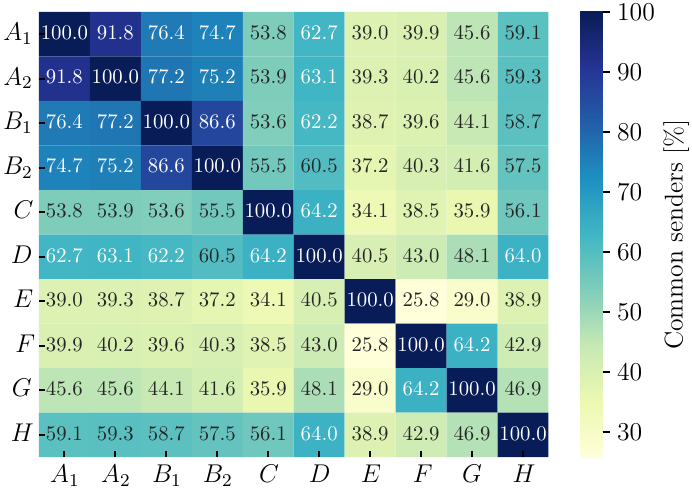}
    \label{fig:heatmap}
}
\hfill
\subfloat[Targeted TCP ports across different sensors from 5 August to 14 August (10 days).]{
\includegraphics[width=0.28\textwidth]{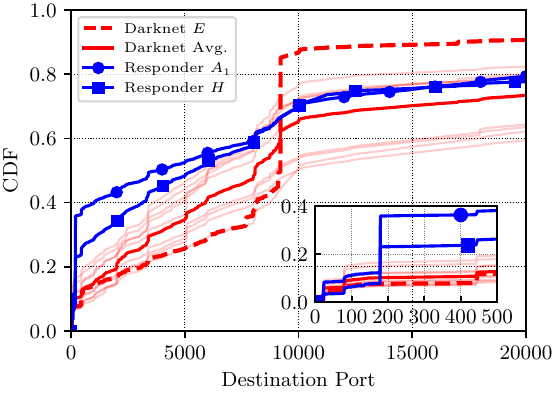}
    \label{fig:cdf_port}
}
\caption{Traffic characteristics observed across distributed darknet sensors.}
\label{fig:combined_analysis}
\end{figure*}

\Cref{fig:timeline_combined} illustrates the temporal evolution of flows received by each IP address across sensors. We define a flow by its standard 5-tuple. For each sensor, we monitor a /24 subnet with only the darknet module enabled (red curves; the solid line denotes the average). On average, each IP observes approximately 10\textsuperscript{4} flows per day, with fluctuations reflecting variations in external sender activity.

The darknet at sensor $E$ emerges as a clear outlier, receiving roughly three to four times more flows than other sensors with a distinct periodic pattern. Manual inspection reveals significantly higher backscatter traffic -- TCP SYN/ACK packets from potential DDoS attack victims where adversaries spoof random source IP addresses, mainly from source ports 53, 80, and 443. Since the $E$ address space resembles private ranges such as 192.168.0.0/16, senders may inadvertently or intentionally select a similar address.

During the period shown in \Cref{fig:timeline_combined}, we activated twice a basic L4 responder in a /28 subnet within sensors $A_1$ and $H$ in August (grey period). They observe an increase in traffic flows, peaking above 10\textsuperscript{5} per day. While similarly configured, the sensors received different traffic volumes, with $A_1$ receiving about twice the traffic (note y-log scale). This confirms that active responders change traffic profiles, and \sysname{} enables observations from multiple, geographically diverse vantage points.

\subsubsection{Common senders on different sensors}

We now focus on differences among senders contributing to unsolicited traffic across sensors. 
Different sensors receive traffic from distinct sender sets; however, overlaps occur. \cref{fig:heatmap} quantifies the ratio of common senders observed by each /24 darknet sensor pair  from 25 July to 3 August. To reduce noise, we focus only on highly active senders, defined as senders generating at least 500 packets over the 10-day observation period. This subset represents approximately 5\% of all observed senders.

Italian sensors $A_1$, $A_2$, $B_1$, $B_2$, $C$, and $D$ share over 50\% of senders, likely because their networks belong to the same Autonomous System (AS), making their address ranges similar to senders. Overlap increases to 86\%–91\% for sensors on contiguous address blocks ($A$ and $B$ cases).

Brazilian sensors $F$ and $G$ share around 64\% of senders, reflecting similar regional visibility within their AS. Interestingly, sensor $H$ (Microsoft Azure East US) shows a stronger affinity with Italian sensors (around 58\% overlap) than Brazilian ones (43\%), suggesting cloud-based sensors attract sender populations more akin to those targeting academic or enterprise networks in Europe rather than Latin America. Sensor $E$, which records the highest unique senders, shares fewer senders with any sensor, confirming its outlier status.

These results demonstrate the added value of a distributed observatory: only by aggregating observations from multiple vantage points can we discern spatial diversity and sensor-specific visibility biases.

\subsection{Destination port across modules}

\cref{fig:cdf_port} shows the cumulative distribution function of TCP SYN traffic per destination port targeted by senders across sensors. Darknet sensors exhibit similar profiles, with traffic concentrated on consistent ports: 22 (SSH), 23 (Telnet), and 2000, commonly associated with brute-force and IoT exploitation. Responder sensors ($A_1$ and $H$) show dissimilar port distributions to darknets and each other: they experience noticeable activity on port 179 (BGP -- see inset). Notably, $A_1$ receives 30\% of packets on port 179, compared to only 10\% for $H$. $H$ sees more traffic on ports 1024$–$10000, consistent with its steeper CDF slope. This confirms that active response modules increase engagement, but different locations yield different results.

Sensor $E$ is a notable exception: its darknet receives over 40\% of TCP SYN traffic on port 9200, typically used by Elasticsearch. This distinct pattern aligns with its anomalous flow volume, suggesting specific automated DDoS campaigns may disproportionately target that address space.

Overall, these findings reinforce the importance of heterogeneous, geographically distributed vantage points: they reveal not only volumetric differences but also distinct behavioural signatures across attackers, networks, and deployment contexts.

\section{Lessons Learned and Conclusion}
\label{sec:conclusion}
We presented \sysname{}, an open, lightweight cybersecurity platform. Leveraging cloud-native technologies and the Infrastructure-as-Code (IaC) paradigm, it provides secure, automated, and resilient orchestration across heterogeneous networks. Our initial multi-sensor deployment demonstrated its scalability and its ability to reveal traffic insights unattainable from a single vantage point. The deployment experience highlighted two critical aspects of operating a distributed infrastructure. First, ensuring cluster resiliency is essential, as sensor failures can disrupt experiments and lead to resource waste. Second, comprehensive experiment automation is necessary to minimise human error and prevent misconfigurations that could compromise the quality of the collected data.

Two main directions will guide our future work. 
First, we plan to expand the platform by introducing a novel set of experiments, such as those involving DNS and certificate announcements, while leveraging this unique and flexible infrastructure to develop and foster AI-driven data analytics capabilities.
Second, we aim to disseminate actionable artefacts to the cybersecurity community, including datasets, periodic reports, blocklist insights, and integrations with existing threat reporting platforms.

\section*{Acknowledgement}
This work was supported by the SERICS (PE00000014) and the ACRE (2022EP2L7H) projects under the MUR National Recovery and Resilience Plan funded by the European Union -- NextGenerationEU. This manuscript reflects only the authors' views and opinions and the Ministry cannot be considered responsible for them.

\bibliography{bibliography}

\section*{Biography}
\vskip -3\baselineskip plus -1fil
\begin{IEEEbiographynophoto}{Andrea Sordello}
(andrea.sordello@polito.it) He is a PhD student at Politecnico di Torino within the SmartData@Polito research center. His research focuses on network traffic collection and analysis.
\end{IEEEbiographynophoto}
\vskip -3\baselineskip plus -1fil
\begin{IEEEbiographynophoto}{Marco Mellia}
[F'21] (marco.mellia@polito.it) He is a full professor at the Politecnico di Torino. His research interests are in Internet monitoring, cybersecurity, and AI applied to different sectors.
\end{IEEEbiographynophoto}
\vskip -3\baselineskip plus -1fil
\begin{IEEEbiographynophoto}{Idilio Drago}
(idilio.drago@unito.it) He is an associate professor at the Università di Torino. His research interests include network security, machine learning, and Internet measurements.
\end{IEEEbiographynophoto}
\vskip -3\baselineskip plus -1fil
\begin{IEEEbiographynophoto} {Rodolfo Valentim} 
(rodolfo.viera@polito.it) is a research assistant specialising in Machine Learning and Cybersecurity at Politecnico di Torino, working on AI solutions for traffic analysis.
\end{IEEEbiographynophoto}
\vskip -3\baselineskip plus -1fil
\begin{IEEEbiographynophoto}{Francesco Musumeci}
[SM'23] (francesco.musumeci@polimi.it) He is an associate professor at Politecnico di Milano, Italy. His research interests are in the fields of ML-aided networking, cybersecurity, converged space-ground networks and disaster resilience.
\end{IEEEbiographynophoto}
\vskip -3\baselineskip plus -1fil
\begin{IEEEbiographynophoto}{Massimo Tornatore}
[F'22] (massimo.tornatore@polimi.it) He is a full professor at Politecnico di Milano, Italy. His research interests include performance evaluation and design of communication networks and machine learning applications for network management.
\end{IEEEbiographynophoto}
\vskip -3\baselineskip plus -1fil
\begin{IEEEbiographynophoto}{Federico Cerutti}
[SM'23] (federico.cerutti@unibs.it) He is a full professor at the Università di Brescia, Italy and visiting professor at Imperial College London, UK.  His research interests are in the areas of cyber threat intelligence and security of AI.
\end{IEEEbiographynophoto}
\vskip -3\baselineskip plus -1fil
\begin{IEEEbiographynophoto}{Martino Trevisan}
(martino.trevisan@dia.units.it) He is an associate professor at the Università di Trieste, Italy. His research interests include network measurements, big data, and cybersecurity.
\end{IEEEbiographynophoto}
\vskip -3\baselineskip plus -1fil
\begin{IEEEbiographynophoto}{Alessio Botta}
(a.botta@unina.it) He is an associate professor at the Universià di Napoli Federico II. His reseach
interests include traffic measurements with applications to cybersecurity, and usage of AI.
\end{IEEEbiographynophoto}
\vskip -3\baselineskip plus -1fil
\begin{IEEEbiographynophoto}{Willen B. Coelho}
(willen@ifes.edu.br) He is currently a PhD student at UFES and IT analyst at the IFES, both in Brazil.
\end{IEEEbiographynophoto}

\vfill

\end{document}